\documentclass[aps,amsmath,amssymb,prd,preprintnumbers]{revtex4}
\usepackage{amsmath}
\usepackage{bm}
\usepackage[dvips]{graphicx}
\newcommand{\be}{\begin{equation}}
\newcommand{\dd}{\displaystyle}
\newcommand{\ee}{\end{equation}}
\newcommand{\bea}{\begin{eqnarray}}
\newcommand{\eea}{\end{eqnarray}}
\newcommand{\de}{\partial}

\begin{document}
\title{Polyakov loop and the color-flavor locked phase of
Quantum Chromodynamics}
\author{M. Ciminale}
\author{G. Nardulli}
\author{M. Ruggieri}
\affiliation{Universit\`a di Bari, I-70126 Bari, Italia}
\affiliation{I.N.F.N., Sezione di Bari, I-70126 Bari, Italia}
\author{R.Gatto}
\affiliation{D\'epart. de Physique Th\'eorique, Universit\'e de
Gen\`eve, CH-1211 Gen\`eve 4, Suisse}
\begin{abstract}
We consider the Polyakov Nambu Jona Lasinio model with three
massless quarks at high density and moderate temperature in the
superconductive color flavor locking phase. We compute the critical
temperature $T_c$ as a function of the baryonic chemical potential
for the phase transition from the superconductive state to the
normal phase. We find that $T_c$ is higher by  a factor 1.5 -2 in
comparison to the model containing no Polyakov loop. We also compute
the specific heat $C_v$ near the second order phase transition and
we show that the inclusion of the Polyakov loop does not change the
value of the critical exponent.

\end{abstract}
 \preprint{BA-TH 570/07}

\maketitle

\section{Introduction}
At small hadronic densities and sufficiently high temperature chiral symmetry is restored and the
nature of the chiral phase transition can be investigated by various effective approaches. One of the
most popular is the Nambu - Jona Lasinio (NJL) model \cite{Nambu:1961tp}, describing the chiral
transition in terms of the $\langle\bar q q\rangle$ order parameter. At high temperatures one also
expects a deconfinement transition \cite{Cabibbo:1979ay}. Its nature is rather clear in pure gauge
theory, because, in absence of quarks, Quantum Chromodynamics at low temperature possesses a $Z_3$
global symmetry, which is spontaneously broken at high temperature $T$. The order parameter for this
phase transition is the Polyakov loop \cite{Polyakovetal} whose expectation value vanishes in the
disordered low temperature phase and is different from zero in the high $T$ phase.

The Polyakov loop is a $SU(3)_c$ matrix in color space given by ($\beta=T^{-1}$)\be L({\bf x})={\cal
P}e^{-i\int_0^\beta dx_4A_4({\bf x},x_4)}\,.\ee For uniform $A_4$ one gets an order parameter that can
be written as follows in the Polyakov gauge\be \Phi\,=\,\frac1{3}
\text{Tr}e^{i\beta(\phi_3\lambda_3+\phi_8\lambda_8)}\,;\label{polyakov}\ee for $T\to\infty$ one has
$\Phi=1$ and in the confined phase $\Phi=0$.

In presence of dynamical quarks there is no clear order parameter
for the deconfinement transition because in this case the $Z_3$
center of the $SU(3)_c$ gauge group is not a good symmetry. Though
one cannot properly speak of a phase transition in this case, the
$T-$dependence of the Polyakov loop can nonetheless be studied by
numerical simulations and one still observes its rise from low to
high temperatures on the lattice.

An interesting and still debated \cite{linking} feature of these
data is that chiral symmetry breaking and the decrease of the
Polyakov loop occur at the same critical temperature
\cite{Fukugitaetal}. It has been argued that a mixing between the
effective models for chiral transition and the Polyakov loop
dynamics  might account for the approximate equality of these
temperatures \cite{Fukushima:2002bk}. A step forward has been
obtained in Refs. \cite{Fukushima:2003fw} and \cite{Meisingeretal},
where  the NJL model is studied in presence of a uniform extended
gauge field $A_4$. Its effect on dynamical quarks is obtained by
identifying the parameters appearing in the Polyakov loop
\eqref{polyakov} with an imaginary quark chemical potential.

This modified NJL model (called Polyakov-NJL=PNJL) is characterized
by a thermodynamical potential $\Omega$ comprising two terms,
$\Omega_{NJL}$ and ${\cal U}(T,\phi)$. $\Omega_{NJL}$ contains the
NJL thermodynamic potential modified by the inclusion of the
imaginary quark chemical potential; in the mean field approximation
it therefore depends on the chiral order parameter and on the
Polyakov loop $\Phi$. ${\cal U}(T,\phi)$ depends only on $\Phi$ and
$T$ and its parameters can be obtained by fitting pure gauge lattice
QCD results. The PNJL model still stands on a conjectural basis. We
do not analyze here its theoretical foundations. Nevertheless we do
attempt to determine some of its possible physical implications.

In \cite{Roessner:2006xn} the PNJL model has been extended to high
baryonic densities for the case of two flavors ($u,d$) by including
a quark chemical potential $\mu$. At moderate $\mu$ and small $T$ a
plausible model describing quark dynamics is the 2SC model
\cite{Rapp:1997zu,Alford:1997zt} characterized by condensation in
the diquark antisymmetric color channel and decoupling of the
strange quark. The two flavor approximation can only be valid at
high, but not very high, densities. At these densities $u$ and $d$
quark masses can play a role, but their effect is included by
considering also condensation in the $\bar q q$ channel.

The aim of the present paper is to consider the case of higher
densities, where all the three light quarks can form color
superconductive pairs. The favored phase for sufficiently high
density is the color flavor locking (CFL) state
\cite{Alford:1998mk}, characterized by three massless quarks,  $qq$
condensation in spin 0, color and flavor antisymmetric state (for
reviews see \cite{reviews}). This result was obtained in a NJL
model, where the gluon interaction is mimicked by a four fermion
interaction and one works in the mean field approximation. The
dominance of the CFL phase  can also be proved in QCD by way of
one-gluon exchange; however this result is valid only at extreme
densities ($\mu\sim 10^8$ GeV) \cite{Rajagopal:2000rs}. These
densities are much larger than those presumably existing in the core
of compact stars, where color superconductivity might be found. For
these latter densities perturbative QCD is of little or no help. On
the other hand the standard non perturbative method, i.e lattice
QCD, is not applicable, as the quark determinant is complex at
$\mu\neq 0$ and MonteCarlo simulations are not directly usable.
Therefore the four-fermion interaction remains as the only practical
way to study the CFL phase.

Though in this approximation gluon interactions are described by an effective four quark interactions,
the Polyakov order parameter can nevertheless play a role, similarly to what happens for two flavors
\cite{Roessner:2006xn}. The study of this role is the aim of this paper, where we present a preliminary
study of the second order phase transition around the critical line $T_c(\mu)$. We will consider only
the case of massless quarks, even though finite mass effects could be included either as free
parameters or by considering condensation in both the diquark and quark-antiquark channels
\cite{Buballa:2003qv}. The reason for this neglect is that the inclusion of, say, the strange quark
mass $M_s$ considerably complicates the analysis, because for $M_s\neq 0$ one should include electric
and color  chemical potentials to enforce electric and color neutrality. We will treat these effects,
as well as the the extension to the gapless CFL (gCFL) phase \cite{Alford:2003fq}, in a future
publication.

Since we neglect mass effects, the free energy depends only on the
order parameter $\Phi$ and the unique gap parameter $\Delta$ (the
role of the gap parameters due the symmetric color channels will be
discussed below). Moreover we are interested only in the transition
line between the CFL and the normal phase in the $T-\mu$ plane.
Therefore we can use a Ginzburg-Landau (GL) expansion near the
critical line. This approximation is discussed in Section
\ref{thermo}. In Section \ref{numerical} we verify that the
transition is continuous and compute the critical temperature as a
function of the quark chemical potential $\mu$. Our result is that
the critical temperature $T_c$ is higher by a factor 1.5 - 2 in
comparison with the treatment of CFL within the original NJL
approximation. We also evaluate the critical exponent $\beta$ that
fixes the relationship between the gap parameter and the temperature
near the phase transition and  we find that including the Polyakov
loop does not change the classical value $\beta=1/2$. The
discontinuity in the specific heat at the second order phase
transition is also evaluated and a comparison of results obtained
with and without Polyakov loop is performed. Finally, some
concluding remarks are contained in Section \ref{conclusions}.

\section{Thermodynamics of the three flavor PNJL model\label{thermo}}
The model we study is described by the quark lagrangian
\begin{equation} {\cal L} = \bar\psi\left(iD_\mu \gamma^\mu +
\mu\gamma_0\right)\psi +{\cal L}_\Delta~.\label{eq:lagr1MU}
\end{equation}
In the above equation we have introduced the coupling of the quarks
to a background temporal gauge field $A_\mu = g\delta_{\mu0}A_\mu^a
T_a$ coupled to the quarks via the covariant derivative $D_\mu =
\partial_\mu - i A_\mu$; $\mu$ is the quark chemical potential.
The term ${\cal L}_\Delta$ is responsible for color condensation. It
 can be obtained in the mean field approximation from a four
fermion interaction term. In the CFL model one has
\begin{equation}
{\cal L}_\Delta = -\frac{\Delta}{2}\left(\psi_{\alpha i}^\dagger
\gamma_5\epsilon^{\alpha\beta I}\epsilon_{i j I} C \psi_{\beta j}^*
+ h.c. \right) - \frac{3\Delta^2}G ~.\label{eq:LagrDelta2}
\end{equation}
Eq.~\eqref{eq:LagrDelta2} describes the fact that in the ground
state one has a non-vanishing expectation value of the di-quark
field operator
\begin{equation}
\langle\psi_{\alpha i} \psi^{\beta j}\rangle \propto \Delta
\epsilon_{\alpha \beta I}\epsilon_{ijI} \neq 0~.\label{eq:SSB}
\end{equation}
The constant $G$ in Eq.~\eqref{eq:LagrDelta2} is the NJL four
fermion coupling constant. In Eq.~\eqref{eq:SSB} we have neglected
the color symmetric channel contribution, as one can prove that it
becomes less and less important when one approaches the second-order
phase transition.

Once the quark lagrangian is specified, the mean field free energy
of the CFL quark matter is easily obtained by integration over the
fermion fields in the generating functional, namely
\begin{equation}
\Omega = {\cal U}(T,\phi) +\frac{3\Delta^2}G
-\frac{T}{2}{\text{Tr}}\sum_{n}\int\frac{d{\bm p}}{(2\pi)^3}
\log\left(\frac{S^{-1}(i\omega_n,{\bm p})}{T}\right)~,\label{eq:due}
\end{equation}
where $\omega_n = \pi T(2n+1)$ are the fermion Matsubara
frequencies, and $S^{-1}$ is the inverse fermion propagator in the
mean field approximation, whose explicit form can be found in
Ref.~\cite{Alford:2003fq}. $S^{-1}$ is in principle a $72\times72$
matrix in color, flavor, spin and Nambu-Gorkov indices. In the high
density limit the effect of the antiparticles can be neglected;
moreover, one can split the left-handed and the right-handed quark
contributions to the free energy, since the quarks are massless and
the condensation does not mix quarks with opposite chirality. Thus
$S^{-1}$ is reduced to a $18\times18$ matrix. It can be rearranged
to a block diagonal form, with a $6\times6$ matrix describing the
propagation of $u_r, d_g, s_b$ quarks, and three $4\times4$ matrices
describing the propagation of $d_r,u_g$, and $s_r,u_b$, and
$s_g,d_b$ quarks. This allows a straightforward extraction of the
quasiparticle dispersion laws, much in the same way as in the
analogous evaluation contained in \cite{Anglani:2006br}, the
difference being that there $\Delta_1=0,\,\Delta_2=\Delta_3$ and
here $\Delta_1=\Delta_2=\Delta_3=\Delta$.

In Eq.~\eqref{eq:due} we have introduced the part of the
thermodynamic potential ${\cal U}(T,\phi) $ which describes the
dynamics of the Polyakov loops in absence of dynamical quarks. In
principle various forms can be used
\cite{Fukushima:2003fw,Roessner:2006xn}; for definiteness we adopt
the form proposed in \cite{Roessner:2006xn}  \be{\cal U}(T,\phi)=
T^4\Big\{-\frac{a(T)}2\Phi^\star\Phi\,+\,b(T)\ln[1-6\Phi^\star\Phi+4(
\Phi^3+\Phi^{\star3})-3(\Phi^\star\Phi)^2]\Big\}\ee where \be
a(T)=a_0+a_1\left(\frac T{T_0}\right)+a_2\left(\frac
T{T_0}\right)\,,~~~ b(T)=b_3
\left(\frac{T_0}T\right)^3\,.\label{rossner}\ee  The Polyakov loop
$\Phi$ can be expressed in terms of one parameter
$\phi\equiv\phi_3$, as the other parameter $\phi_8$ can be always be
absorbed by a redefinition of $\phi_3$. It is given by \be
\Phi=\Phi^\star=\frac{1+2\cos(\beta\phi)}3\,.\ee Numerical values of
the coefficients have been fitted in \cite{Roessner:2006xn} using
lattice data \cite{Boyd:1996bx}:\be
a_0=3.51,\,a_1=-2.47,\,a_2=15.2,\,b_3=-1.75\,,\ee together with the
deconfinement temperature $T_0=270$ {\rm MeV}\,. The use of a
definite form for ${\cal U}(T,\phi)$ is not a limit of our
computation because other functional dependences produce similar
results as they are derived from the same lattice data set. The NJL
coupling $G$ should in principle depend on $\Phi$ because the four
fermion coupling is induced by gluon dynamics. However following
\cite{Fukushima:2002bk} we will neglect this effect. In final
results we will trade $G$ for the value $\Delta_0$ of the CFL gap at
$T=0$ using the weak coupling formula~\cite{Schafer:1999fe} \be
\frac{3}G= \frac{6\mu^2}{\pi^2}\ln\frac{2\delta}{\dd
2^{\frac13}\Delta_0}~. \label{eq:Weak}\ee In the above relation
$\delta$ is an ultraviolet cutoff, introduced to ensure ultraviolet
convergence of the loop integrals. By means of Eq.~\eqref{eq:Weak}
the coupling strength is represented by the parameter $\Delta_0$.

Performing the summation over the Matsubara frequencies in
Eq.~\eqref{eq:due} one gets\bea\Omega&=&{\cal
U}(T,\phi)+\frac{3\Delta^2}G-2\sum_{i=1}^9\int
\frac{d^3p}{(2\pi)^3}\Big( T\ln\Big|1+e^{-\beta\epsilon_j}\Big|+
\Re(\epsilon_j-p+\mu)\,+\,\cr &&~~~~~~~~~~+
T\ln\Big|1+e^{-\beta\tilde\epsilon_j}\Big|+\Re(\tilde\epsilon_j-p-\mu)
 \Big)~.\label{uno}\eea
In Eq.~\eqref{uno} $\epsilon_j$ are the energies of quasiparticles
and $\tilde\epsilon_j$ are obtained from $\epsilon_j$ by the
substitution $\xi=p-\mu\to p+\mu$. The terms with $\tilde\epsilon_j$
correspond to antiparticles. They are put here for completeness but
omitted in numerical evaluations. The dispersion laws for the nine
quasiparticles can be derived by the standard methods. Since the
resulting expressions are cumbersome and our study is limited to the
critical line we present here their expression only for small values
of the gap parameters, ie in the GL approximation. One obtains\bea
\epsilon_1&=&\epsilon_2^\star=
\left(\xi+i\phi+\frac{\Delta^2}{2\xi}\frac{8\xi^2+\phi^2}{4
\xi^2+\phi^2}\,- \,i\frac{\Delta^2\phi}{4\xi^2+\phi^2} \right)\cr
\epsilon_3&=&\xi\left(1+\frac{4\Delta^2}{4\xi^2+\phi^2} \right)\cr
\epsilon_4&=&\epsilon_5^\star=\xi+i\phi+\frac{\Delta^2}{2\xi}\cr
 \epsilon_6&=&\epsilon_8^\star=\xi+i\phi\,+\,\frac{\Delta^2}{4\xi^2+\phi^2}
 \left(2\xi-i\phi
\right)\cr \epsilon_7&=&\epsilon_9^\star=
\xi+\,\frac{\Delta^2}{4\xi^2+\phi^2}
 \left(2\xi-i\phi
\right)\label{unobis}\ .\eea We have also computed the coefficients
of the ${\cal O}(\Delta^4)$ term  but we do not report them here
(they are needed to control that the phase transition is continuous
at $T_c$ and to compute the gap, see below).

The gap parameter $\Delta$ and the background gauge field $\phi$ at
a fixed temperature and chemical potential are obtained solving the
equations \bea \frac{\de\Omega}{\de\phi}&=&0\,,\label{prima}\\
\frac{\de\Omega}{\de\Delta}&=&0\,.\label{prima2}\eea  Near the
critical temperature $T_c$ one can expand $\Omega$ in
Eq.~\eqref{uno} as
follows\be\Omega(\Delta,\phi)-\Omega(0,\phi)\sim\frac\alpha2\Delta^2+
\frac\beta4\Delta^4~. \label{eq:GL}\ee
The critical temperature
$T_c$ is obtained at a fixed $\mu$ as in the usual BCS theory by
solving the equation $\alpha(T_c)=0$, with $\alpha$ given by
\begin{eqnarray}
\alpha&=&
\frac{6}{G}\left[1+~GT~\frac{2}{3}\frac{\mu^2}{\pi^2}\sum_{n}
\int_{-\delta}^{~\delta}\!d\xi~
\frac{\left[3\cdot(l_0^2-\xi^2)^2+\phi^2\cdot(l_0^2+3\xi^2)\right]}{(l_0^2-\xi^2)\left[(l_0+\xi)^2+\phi^2\right]
\left[(l_0-\xi)^2+\phi^2\right]}\Big|_{l_0=i\omega_n}\right]\nonumber
\\
&&~~~~~=~\frac{12\mu^2}{\pi^2}\left(\ln\frac{2\delta}{\dd
2^{\frac13}\Delta_0}\,+\,\frac13\int _{-\delta}^{+\delta}d\xi
f(\xi,\phi)\right) \label{eq:alpha}
\end{eqnarray}
and $f$ defined by
\begin{equation}
f(\xi,\phi)=-\, \frac {2\xi}{4\xi^2+\phi^2}
\tanh\frac{\beta\xi}2\,-\,2\Re\frac{4\xi-i\phi}{4\xi(2\xi-i\phi)}
\tanh\frac{\beta(\xi-i\phi)}2~.
\end{equation}This expression for $\alpha$ is identical to the result
obtained by \eqref{uno} using the dispersion laws \eqref{unobis} up
to $\Delta^2$. On the other hand the the coefficient $\beta$ is
given by \be\beta = T \frac{\mu^2}{2\pi^2}\sum_{n}
\int_{-\delta}^{~\delta}\!d\xi~ \frac{8\Delta^4 \cdot{\cal
F}(l_0,\xi,\Phi)
}{\left\{(l_0^2-\xi^2)\left[(l_0+\xi)^2+\phi^2\right]
\left[(l_0-\xi)^2+\phi^2\right]\right\}^2}\Big|_{l_0=i\omega_n}~,\ee
where \bea {\cal F}(l_0,\xi,\Phi)&=& 6 l_0^8  -
l_0^6\cdot(24\xi^2+5\phi^2)+ l_0^4\cdot(36\xi^4 + 21\xi^2\phi^2
-4\phi^4)\nonumber\\&&-l_0^2\cdot(24\xi^6+26\xi^4\phi^2-24\xi^2\phi^4+\phi^6)+
\xi^2\cdot(6\xi^2 -\phi^2)\cdot(\xi^2 +\phi^2)^2~.\eea The summation
over Matsubara frequencies in the expression of $\beta$  can be
performed analytically, but the final expression is involved and we
omit it for simplicity.

\section{Numerical results\label{numerical}}

To get the critical temperature $T_c$ we solve the equation
$\alpha(T_c) = 0$ with $\alpha$ given by Eq.~\eqref{eq:alpha} and
$\phi$ obtained by Eq. \eqref{prima}. We have checked that the phase
transition is of the second order since $\beta(T_c)
> 0$.
It is well known that in the case $\phi = 0$ one has in the CFL
phase $T_c/\Delta_0 \simeq 0.71$, see for
example~\cite{Fukushima:2004zq}. On the other hand in the 2SC phase
one has $T_c/\Delta_0 \simeq 0.57$ as in ordinary BCS
superconductors. This difference is related to the fact that in the
CFL model one has eight gapped modes with gap $\Delta$ and one mode
with gap $2\Delta$.

\begin{figure}[ht]
\begin{center}
\includegraphics[width=9cm]{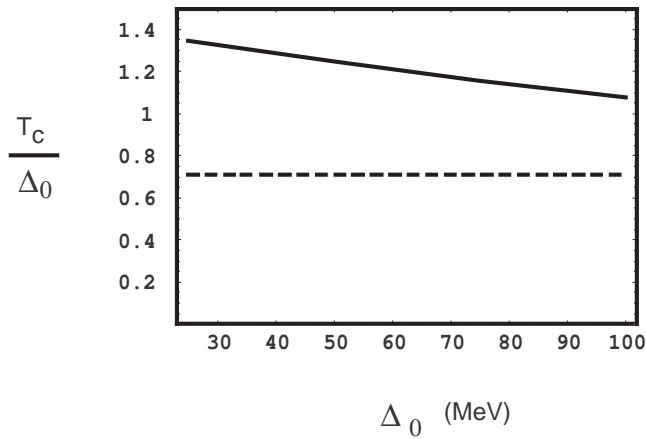}~~~~~~~~~~~~~~~~~~~~~~~~~~~
\end{center}
\caption{\label{Fig:Fig1} Ratio $T_c/\Delta_0$ against $\Delta_0$
(MeV), with (solid line) and without (dashed line) Polyakov loop, at
$\mu = 500$ MeV.}
\end{figure}

The result of the numerical evaluation of $T_c$ is shown in
Fig.~\ref{Fig:Fig1}, where we plot the ratio $T_c/\Delta_0$ with
(solid line) and without (dashed line) Polyakov loop, at the
reference value $\mu = 500$ MeV.
We notice that introducing self-consistently  the parameter $\phi$
implies a significant increase of the critical temperature. This
effect has been noticed also in the two flavor
model~\cite{Roessner:2006xn}.

Next we turn to the behavior of the gap parameter $\Delta$ for
temperature close to $T_c$. We find
\begin{equation}
\frac{\Delta(T)}{T_c} = k(\Delta_0) \left(1 -
\frac{T}{T_c}\right)^\beta~,~~~~~T\rightarrow T_c^-~,
\end{equation}with $\beta=1/2$. The value of the critical exponent
is the same as in BCS superconductors. However the presence of the
Polyakov loop affects the constant $k$ in two ways. First, it gives
it a dependence on $\Delta_0$ that is absent in the BCS and in the
two flavor color superconductor. Second, it changes its numerical
values. For example for the 2SC case $k\simeq 3.1$; in the present
case $k=1.7$ and $2.2$ for $\Delta_0=40$ and 100 MeV respectively.

\begin{figure}[ht]
\begin{center}
\includegraphics[width=8cm]{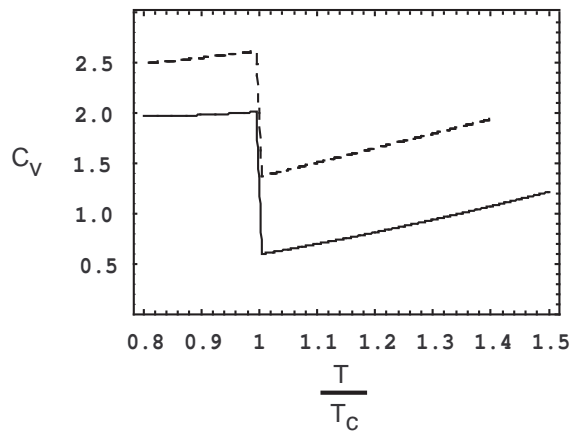}~~~~~~~~~~~~~~~
\end{center}
\caption{\label{Fig:Fig1bis} Specific heat $C_v$ (Units: 10$^7$
MeV$^3$) against $T/T_c$, with (solid line) and without (dashed
line) the Polyakov loop,  for $\mu = 500$ MeV and $\Delta_0 = 25$
MeV.}
\end{figure}

The knowledge of $\Delta(T)$ near $T_c$ allows to determine some
thermal properties of the model. For example we compute the specific
heat as a function of the temperature, near $T_c$\,. It is given by
\begin{equation}
C_v = -T \frac{\partial^2 \Omega(\Delta,\phi)}{\partial T^2}~.
\label{eq:O}
\end{equation}

We show the result of this calculation in Fig.~\ref{Fig:Fig1bis},
with (solid line) and without (dashed line) Polyakov loop, for  $\mu
= 500$ MeV and $\Delta_0=25$ MeV (for other values of $\Delta_0$ we
find qualitatively similar results). We notice that including the
Polyakov loop slightly decreases the specific heat and increases a
bit its discontinuity around $T_c$.

\section{Conclusions\label{conclusions}}
In this paper we have studied the effect of the inclusion of the
Polyakov loop on the NJL description of the CFL model. We have
restricted our attention to a temperature range close to the
critical temperature of the second order phase transition. We have
found that introducing the Polyakov loop significantly increases the
critical temperature, the effect being more important in the weak
coupling regime. This increase  may have some phenomenological
consequences, both for astrophysical systems and for future
experiments at GSI, if the proposed facility SIS100/200 \cite{GSI}
will be able to reach the hadronic densities needed for color
superconductivity. Needless to say, one has to stress the heuristic
use of the Polyakov loop when quarks are dynamical. Already their
presence destroys the center symmetry of pure gauge QCD. More
theoretical investigation will be needed on the PNJL model to
ascertain its possible regions of validity. Nevertheless we felt
that it is useful to investigate the effect of the Polyakov loop in
some portions of the QCD phase diagrams where a direct QCD treatment
is not available at the present.

 We have studied the behavior of the gap
parameter $\Delta(T)$ for $T\approx T_c$, showing that the Polyakov
loop does not modify the critical exponent $\beta = 1/2$, but only
the pre-factor. In ordinary superconductor the pre-factor does not
depend on $\Delta_0$; on the other hand, the presence of the
Polyakov loop results in a pre-factor dependent on the strength of
the coupling.

A quantity of interest is the specific heat $C_v$ since it can be
measured experimentally. At the second order phase transition $C_v$
is discontinuous, in the superconductive phase being larger than in
the normal phase. Although the effect of the Polyakov loop is to
decrease the absolute value of $C_v$, the discontinuity $\Delta C_v$
with $\phi\neq 0$ is larger than the corresponding value at $\phi =
0$.

Further developments include the treatment of the strange quark
mass, as well as the study of the thermodynamics of the CFL
superconductor with Polyakov loop at lower temperatures. While the
effect of the Polyakov loop is not expected to modify strongly the
thermodynamics at small $T$, it is known that the finite strange
quark mass affects the phase diagram of QCD and its role in the
present model may be of interest as well.

\end{document}